\documentclass[twocolumn,prb,color,psfig,showkeys,superscriptaddress]{revtex4}

\usepackage{graphicx}
\usepackage{epsfig}
\usepackage{amsmath}
\usepackage{amssymb}

\newcommand{\eps}{\ensuremath{\varepsilon}}
\newcommand{\s}{\ensuremath{\sigma}}
\newcommand{\w}{\ensuremath{\omega}}
\newcommand{\p}[1]{#1^{\prime}}

\begin{document}

\title{Superfluid -- Bose glass transition in two dimensions \\ at $T = 0$:
analytic solution of a mode-coupling toy model.}

\author{E.V. Zenkov}
\email{eugene.zenkov@mail.ru}
\affiliation{Ural State Technical University, 620002 Ekaterinburg, Russia}


\begin{abstract}
Analytic expression for the memory function and the  optical conductivity of the two-dimensional Bose gas with logarithmic interaction at $T = 0$ in presence of point-like
impurities is obtained within the mode-coupling approximation. Depending on the
value of a dimensionless combination of the model parameters proportional to the strength of the impurity potential, two different phases are distinguished, viz. the disordered superfluid and insulator (Bose glass), separated by an intermediate (quasi)metal state.
\end{abstract}

\maketitle

\section{Introduction}

The behaviour of many-particle classical and quantum systems in presence of a random
perturbation is a long-standing problem of great scientific interest.


Various methods have been developed to describe the behaviour of a system subject to the
random external force, such as the coherent potential and other effective
medium types of approximations \cite{krum, hori},
the self-consistent hydrodynamic scheme for the description of the
overdamped charge dynamics in presence of a quenched disorder, elaborated by W.
G\"otze and coworkers \cite{GotzePhil}, the replica theories of flux lattices and Wigner crystals \cite{giamarchi}.

It is interesting to observe, that for very different model systems in the
localization regime, many quantities of physical interest, such as the dynamic
conductivity, were found to manifest an intimate resemblance. E.g., in the
three-dimensional (3D) electron system, the Drude tail of the optical
conductivity flattens out and gradually evolves into a finite energy peak as the
system approaches the Anderson transition \cite{GoldFermi} from the metallic side of
the phase diagram. Similar results are obtained for the two-dimensional (2D) electron
gas and also for 2D and 3D Bose gas at zero temperature \cite{GoldHTS, GoldBose},
where the effects of the disorder lead to the decay of the coherent response of the
condensate $\s_{coh}$ and the transfer of the spectral weight to a finite-energy broad
feature. The same general picture holds also for pinned density waves \cite{gruner}
and Wigner crystals \cite{giamarchi}, where the conductivity displays the
inhomogeneously broadened feature at the pinning energy.

However, the understanding of the disorder effects in complex quantum systems is still an issue, that have been given the new impetus since  the discovery of the strongly correlated oxides.
An extreme complexity of these problems usually rules out the possibility of a
tractable analytic treatment.
In this article we present a simple exactly solvable model of the superfluid-insulator transition in the 2D Bose gas with logarithmic interaction in presence of the random short-range potential of the point defects at zero temperature.

The key result of our analysis is the analytic expression for the memory function,
obtained within the framework of the mode-coupling theory \cite{GoldBose}. This
enables us to explore the evolution of the optical conductivity in the full range of
the relevant physical parameters and to trace the superfluid-insulator transition,
that takes place, when a dimensionless parameter of the disorder exceeds a critical
value. Despite a number of simplifying assumptions, we believe, that the theory,
developed in the next section, is of a certain interest, being an elementary but
instructive exactly solvable model of the disorder-induced glass transition.

\section{The model}

The central role in our analysis plays the memory function
$M(z)=M^{\prime} + i M^{\prime\prime}$, that measures the relaxation rate of the
current excitations and incorporates the effects of an external random perturbation
due to the impurities.

Within the mode coupling theory \cite{GotzePhil} the memory function obeys the
self-consistency relation:
\begin{widetext}
\begin{equation}\label{memory}
 M(z) = i \gamma + \frac{1}{2 n m} \int \frac{\Phi_0(k, z + M(z))}{1
 + M(z) \Phi_0(k, z + M(z))/\chi(k, z=0)} k^2 \langle |U(k)|^2
 \rangle d\mathbf{k},
\end{equation}
\end{widetext}
where $\gamma$ is a bare relaxation constant, corresponding to some internal current
decay channels, $U(k)$ is the Fourier transform of the random potential, $n$ and $m$ are the concentration and the mass of the particles, respectively. The angular
brackets stand for the averaging over the disorder. The density response function $\Phi_0$ is related to the generalized susceptibility $\chi_0(k, z)$ of the system without disorder:
\begin{equation}\label{fi2}
 \Phi_0(k, z) = \frac{\chi_0(k, z) - \chi_0(k, 0)}{z}.
\end{equation}
Thus, the free model being specified by its correlation functions, eq. (\ref{memory})
offers a direct way to explore the effects of the disorder on its electric and optical
properties.

We consider the 2D Bose gas at zero temperature, where all particles are
condensed into the superfluid phase. Hence, no bare mechanisms of relaxation are
presumed \footnote{The account of nonzero bare damping does not change the obtained
results in a qualitative way, but merely broadens the incoherent wing (see
below).} and $\gamma = 0$ in (\ref{memory}).
The susceptibility of noninteracting 2D bosons is readily obtained \cite{hines}:
\begin{equation}
 \chi_{nonint.}(k, z) =  \frac{2  n  \eps_k}{\eps_k^2  -  z^2},
\end{equation}
where $\eps_k = k^2/2 m$ is the dispersion of free particles.
The interparticle interaction $V$ modifies the susceptibility. A qualitative estimate of this effect may be done within the random phase approximation:
\begin{equation}\label{rpa}
 \chi_0(k, z) =  \frac{\chi_{nonint.}(k, z)}{1 + V(k) \chi_{nonint.}(k, z)}.
\end{equation}
Having in mind to work out a simple tractable model, we neglect here a certain
limitation of this mean-field type of approximation in low dimensions, as well as the
local-field corrections \cite{zeitschrift}. On the other hand, the mean-field concept
retains its significance as a highly intuitive approach, that makes it possible to get
insight even into the problems, where its formal substantiation is problematic, e.g.
in the physics of the superfluid films \cite{ambegaokar} and the plane vortices in
quasi-two-dimensional magnets \cite{huber}.

In purely two-dimensional systems, such as the planar vortices, the axially
symmetric solution of the 2D Laplace equation provides the following form of
the interparticle interaction:
\begin{equation}\label{log}
  V(r) = Q^2 \ln\left(\frac{r_0}{r}\right),
\end{equation}
(the two-dimensional analog of the Coulomb potential), where Q is the boson charge and
$r_0$ is an appropriate length scale, that may be considered as a mean interparticle
distance, $r_0 \sim n^{1/2}$, where $n$ is the two-dimensional concentration of
particles. Thus, the potential $V(r)$ is repulsive for $r < r_0$ and attractive for $r
> r_0$, favouring the homogenisation of the system. The ground-state properties of the free 2D Bose gas with the interaction (\ref{log}) have been investigated in \cite{tanatar}. Correspondingly, the potential in momentum space is:
\begin{equation}\label{logK}
 V(k) = \frac{2\,\pi\,Q^2}{k^2},
\end{equation}
that may be obtained as the 2D Fourier transform of $V(r) \exp(- \alpha\,k\,r)$ in the limit $\alpha \rightarrow 0$.

Hereafter it is convenient to introduce the dimensionless variables
$\widetilde{p} = p/p_0$,
$\widetilde{\w} = \w/\eps_0$,
$\widetilde{M} = M/\eps_0$,
defining the units of momentum and energy as
$p_0 = (8 \pi Q^2 \hbar^2 m n)^{1/4}$ and $\eps_0 = p_0^2/2 m$, respectively
\footnote{Note, that in 2D the dimensionality of charge ($Q$) differs from that of the
conventional 3D charge by a factor $length^{-1/2}$}. In what follows, we shall
consider the case of identical impurities, with random coordinate distribution. The
average ($\langle \dots\rangle$) in rhs. of eq. (\ref{memory}) then reduces to the
multiplication by the impurity concentration $n_i$. Substituting eqs. (\ref{fi2}-\ref{rpa}) with (\ref{logK}) into (\ref{memory}) yields the nonlinear equation on the memory function, where all quantities are reduced to the dimensionless form and tilda's are omitted:
\begin{equation}\label{Mlog1}
  M = \frac{8 \Lambda}{\pi} \int \limits_{0}^{\infty}
  \frac{(z + M) k^5 \varphi(k)^2}{(1 + k^4) \left(1 + k^4 - z (z + M)\right)} dk.
\end{equation}
Thus, the optical response of the model is governed by a single parameter
\begin{equation}\label{lamda}
 \Lambda = \frac{\pi}{2} \frac{n_i}{n} \left(\frac{U_0}{\eps_0}\right)^2 =
         \frac{n_i m}{4 n^2}\left(\frac{U_0}{\hbar  Q}\right)^2,
\end{equation}
where $U_0$ measures the strength of the
impurity potential, $U(k) = U_0 \varphi(k)$. Note, that $U_0$ has the dimension of
energy, and that of $\varphi(k)$ is $1/k^2$. The similar non-linear equations
on $M(z)$ have been derived early on for the 2D and 3D Bose gas with the Coulomb
repulsion \cite{GoldHTS, GoldBose}. However, their analytic solution were never
possible because of the complexity of the integrand.

In what follows, we focus on the special case of the short-range impurity centers,
that is usually considered in the theory of pinned density waves and Wigner crystals:
$\varphi(r) = \delta(r)$, whence $\varphi(k) = 1$. Then, the integral in rhs. of
(\ref{Mlog1}) can be solved to yield the quadratic equation for $M$,
whence we obtain the explicit expression for the memory function:
\begin{equation}\label{Mlog}
  M = 2 \Lambda \Biggl(\frac{1 - \Lambda}{z} -
   \sqrt{\left(\frac{1 - \Lambda}{z}\right)^2 - 1}  \Biggr),
\end{equation}
where the root with the non-negative imaginary part is to be chosen. This is the main
result of the present article. Note, that within the mode-coupling theory eq.
(\ref{Mlog}) is an {\it exact} relation.

\begin{figure}[t]\label{fig1}
\includegraphics[width=\linewidth]{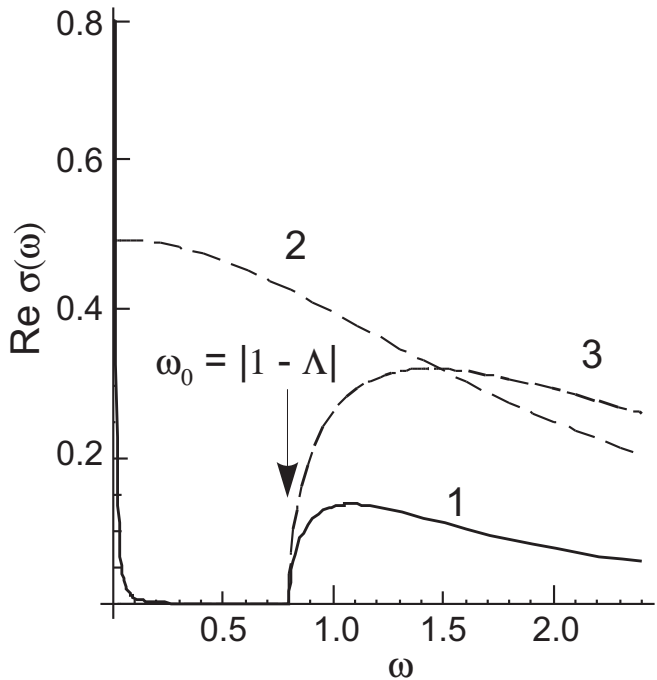}
\caption{The optical conductivity spectra of 2D Bose gas with point-like
impurities at different levels of the disorder: 1 - $\Lambda = 0.2$
(superfluid), 2 - $\Lambda = 1.0$ (metal), 3 - $\Lambda = 1.8$ (insulator). The
arrow indicates the edge of the incoherent wing, that falls to the same frequency
for both curves 1 and 3. For the sake of
clearness, the infinitely narrow $\delta$-like peak in the superfluid phase is
shown broadened.}
\end{figure}

\begin{figure}[t]\label{fig1}
\includegraphics[width=\linewidth]{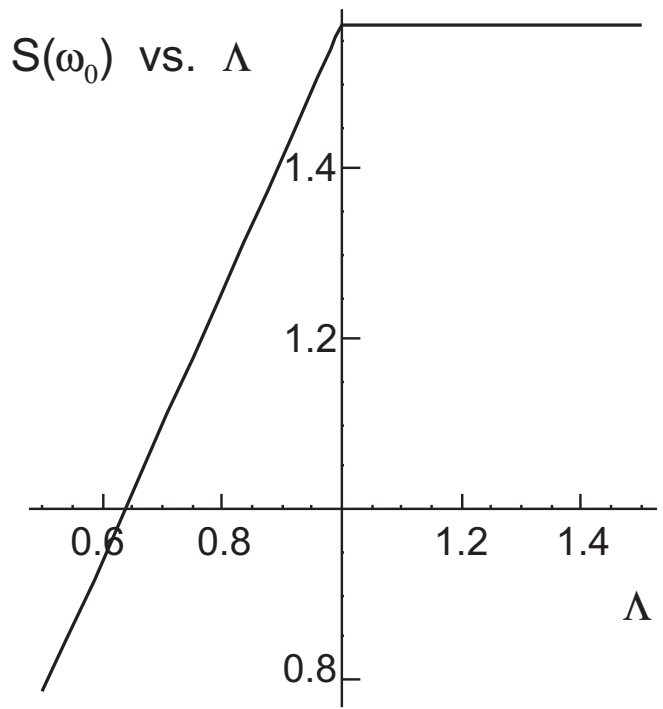}
\caption{The dependence of the spectral weight $S$ of the incoherent wing, eq.(\ref{s2}), on
the disorder parameter $\Lambda$. The knee at $\Lambda_c = 1$ indicates the superfluid-insulator transition.}
\label{fig2}
\end{figure}

\section{Optical conductivity and the superfluid - glass transition}

The optical response of the system is most conveniently studied in terms of the
optical conductivity. Within the memory function approach this is expressed in the
generalized Drude form:
\begin{equation}\label{drude}
   \s^{\prime} + i \s^{\prime\prime}  = i \frac{\w_p^2}{z + M(z)},
\end{equation}
where $z = \w + 0 i$ is the complex frequency, $\w_p$ is the plasma
frequency ($\w_p^2 = Q^2 n/m$, where $n$, $Q$ and $m$ are the concentration, charge and mass of bosons, respectively).

The explicit analytic expression for the memory function (\ref{Mlog}) makes it
possible to explore various regimes of the conductivity of the system. First of all,
it is straightforward to show, that this results preserves the sum-rule:
\begin{equation}\label{sumrule}
 \int\limits_0^{\infty} \p{\s}(\w) d\w = \frac{1}{2} \pi \w_p^2,
\end{equation}
that we notice as an important sign of the sound behaviour of the present model.

Turning to the detailed analysis of the conductivity spectrum, let us consider first
the weak disorder limit, $\Lambda \ll 1$. Expanding the generalized Drude formula
(\ref{drude}) to first order in $\Lambda$, we arrive at the following expression for the
real part of the optical conductivity at finite frequencies:
\begin{equation}\label{incoh}
  \p{\s}(\w)_{incoh} =
   \begin{cases}
     2 \Lambda  \displaystyle \frac{\sqrt{\w^2 - 1}}{\w^3}, & \w \geqslant 1 \\
     0, & \w < 1.
   \end{cases}
\end{equation}
Thus, the spectrum shows a gap, that a more accurate treatment finds to extend at
$\w_0 = |1 - \Lambda|$. The free system parameters being fixed, this gap frequency is
governed by the amplitude of the disorder (pinning) potential. Above the gap, the
optical conductivity $\p{\s}_{incoh}$ displays a broad asymmetric feature, shown in
fig. \ref{fig1}. It corresponds to the localized plasma oscillations of the charge
density and is hereafter called the incoherent wing. Such a behaviour is typical of the
charge density waves.
Interestingly, we recover in (\ref{incoh}) the expression
for the conductivity of pinned CDW, first derived by Fukuyama and Lee \cite{fukuyama}.
We believe this result to be the consequence of a nontrivial analogy between the
pinned CDW and the disordered Bose condensate, albeit such a literal coincidence of
the formulas is rather accidental. The point is that pinned CDW is a collective state,
that can be described by a model of a particle in a potential well \cite{fisher}. On
the other hand, the condensate at $T = 0$ is also a collective (coherent) state, that
behaves as a single particle.

However, it follows from the sum-rule (\ref{sumrule}), that the contribution of
$\s_{incoh}$ (\ref{incoh}) to the optical conductivity spectrum cannot be the only one,
since it vanishes as $\Lambda$ tends to zero. As eq.(\ref{incoh}) is reasonable at
$\w$ large, the careful examination of the conductivity at low frequencies is
required. The corresponding expansion of the memory function (\ref{Mlog}) is:
\begin{equation}\label{Mz0}
  M(z) \simeq  \left(1 - sign(1 - \Lambda)\right) \frac{\Lambda (1 - \Lambda)}{z} +
\frac{\Lambda}{1 - \Lambda} sign(1 - \Lambda) z.
\end{equation}

The behaviour of the system depends critically on the magnitude of $\Lambda$. According to eq.
(\ref{Mz0}), the memory function is linear near the zero frequency for small disorder parameter, ($\Lambda < 1$). The corresponding optical conductivity
is:
\begin{equation}\label{coh}
\s(\w)_{coh} = (1 - \Lambda) \frac{i}{z}\Bigl|_{z = \w + 0 i}.
\end{equation}
This expression describes the coherent response of the condensate. Its spectral weight
drops linearly as the disorder parameter $\Lambda$ increases. According to the
sum-rule (\ref{sumrule}), this implies the emergence of the new spectral feature at
higher frequencies, described by $\s_{incoh}$ (\ref{incoh}). The full conductivity
spectrum is the sum of the coherent (\ref{coh}) and incoherent (\ref{incoh}) contributions,
as depicted in fig. \ref{fig1}.

In the strong disorder regime ($\Lambda > 1$), one
obtains from (\ref{Mz0}) in the leading order $M \sim  -1/z$, and $\p{\s}(0) = 0$. Within the theory of the liquid-glass
transitions \cite{liquids} this pole of the memory function is known as the
non-ergodicity pole and its appearance is identified with the onset of the glass phase
as it calls forth the long-lived density fluctuations $\langle \delta \rho(k,
t) \delta \rho(k, 0) \rangle$, that do not decay to zero at large times. Thus, we
call the insulating phase of our model the Bose glass.

The superfluid-glass transition may be clearly visualized by considering
the partial spectral weight of the incoherent contribution,
\begin{equation}\label{s2}
 S_{incoh}  =  \int_{|1 - \Lambda|}^{\infty}  \p{\s}(\w)  d\w.
\end{equation}
This quantity is shown in fig.\ref{fig2} against the disorder parameter $\Lambda$. It can be
seen, that $S$ grows progressively as $\Lambda$ approaches unity from below, i.e. the
spectral weight moves away from the condensate to the localized modes. However, it does no
longer change as soon as $\Lambda > 1$, although $\w_0$ continues to grow. This is
because no spectral weight is now contained below $\w_0$ and the superfluidity is
completely suppressed by disorder.

The system undergoes the superfluid-insulator transition via an intermediate state,
that may be called metallic. The high-frequency asymptotics of the optical
conductivity in both phases follows the Drude law with the damping parameter equal to
$2 \Lambda$, as can be seen from the limiting form of (\ref{Mlog}). At the
critical transition point, $\Lambda = 1$, the spectral weight of the condensate
vanishes, and so does the gap frequency $\w_0$, while the memory function is $M =
2 i$. This obviously corresponds to the conventional Drude optical conductivity with
the dispersionless relaxation rate in the full spectral range.

\section{Summary and conclusions}

In this paper we have studied analytically the superfluid-glass transition within the
purely two-dimensional Bose gas with logarithmic interparticle interaction at $T = 0$
K with static disorder, represented by the point-like random impurities. The
dielectric properties of the system are considered proceeding from the explicit
expression for the memory function $M(z)$, obtained within the mode-coupling theory.
The onset of the glass phase manifests itself as the appearance of the nonergodicity
pole \cite{liquids} of the memory function, $M(z) \sim 1/z$.

The above analysis shows, that the phase diagram of the model is delimited by the
following combination of the physical parameters:
\begin{equation}
 \Lambda_c = \frac{n_i m}{4 n^2}\left(\frac{U_0}{\hbar  Q}\right)^2 = 1,
\end{equation}
Untill $\Lambda < 1$, the optical conductivity $\s(\w)$ spectrum consists of the response of the condensate $\propto \delta(\w)$, where $\delta(w)$ is the Dirac delta function, and an inhomogeneously broadened feature at $\w \geq |1 - \Lambda|$. The latter one is typical to the pinned density waves and falls off as $1/\w^2$ at large frequencies.

At $\Lambda > 1$ the Dirac delta function contribution disappears and the system
undergoes the transition to a localized phase. Hence, large value of either the boson
effective mass $m$, the amplitude of the impurity potential $U_0$ or the concentration
of impurities $n_i$, other parameters being fixed, favours the localization and
suppresses the superfluid phase, while strong boson interparticle repulsion $Q$ and
large boson concentration $n$ have an opposite effect.

A natural question arises on the internal structure of the novel localized phase.
Within the present hydrodynamic approach this problem cannot be resolved on the
microscopic level. Qualitatively, the phase transition may be realized as a condensate
fragmentation in the rugged potential landscape, that entails the breakdown of the
long range non-diagonal order. The system breaks up into the patches, trapped by the
randomly distributed pinning centers. By analogy with the mode-coupling theory of
simple liquids, where the similar transition is referred to as the liquid-glass
transition, we call the localized phase of the present model the Bose-glass phase (see also \cite{GoldBose, zeitschrift}, where this name was first introduced).

Now, let us dwell in more details on the comparison of our results with some other approaches to the localization and superfluid-insulator transition in a system of 2D bosons.
In Ref.\cite{das} the phase diagram of the disordered 2D Bose liquid is discussed in terms of two dimensionless parameters, $J/V$ and $\Delta/V$, where $J$ is the zero point energy of the bosons, $V$ measures their short-range repulsion and $\Delta$ is a disorder parameter. It was found, that large value of $J/V$ favours the superfluid (SF) phase, while the decreas of $J/V$ or the increase of $\Delta/V$ leads to the localization and the supression of SF phase.
In our case the only parameter $\Lambda$, eq.(\ref{lamda}), governs the phase diagram.
The ratio $n^2/n_i$, that enters therein may be understood as the concentration of bosons per a patch around an impurity. Thus, the first factor in $\Lambda$ is proportional to the inverse energy per boson in such a patch \cite{schick}, $n_i m/(\hbar^2 n^2) \sim J^{-1}$.
Hence, $\Lambda$ may be written in terms of Ref.\cite{das} as: $\Lambda \sim \Delta^2 /(J\,V^2)$. It can be seen, that the increase of $\Lambda$, that describes the gradual supression of SF phase and the onset of the localization, qualitatively agrees with the mentioned general results of Ref.\cite{das}.

The problem of the relevance of the model to the realistic physical
systems is subtle, since the real particles, either boson or fermions, are likely to
interact via not logarithmic, but rather the screened Coulomb potential.
However, the effective model of 2D Bose liquid with the logarithmic interaction at $T = 0$ is related to the physics of the pinned vortices in the bulk type II
superconductors \cite{nelson}.
We do not scrutinize this question any further in the present article.

We believe the model considered to be of importance on its own
right, as it exhibits a rather non-trivial behaviour and allows a fully analytic
treatment. The comparison of our results with those, obtained for the other model of the disordered Bose systems numerically and semi-analytically within the same theoretical framework
\cite{GoldBose, GoldHTS}, leads to the conclusion, that in both cases the phase
diagram and the optical spectra are essentially the same and are fairly
weakly sensitive to the character of the interparticle interaction, the impurity
potential and the dimensionality of the systems. Thus, we hope the present study may
shed light on the behaviour of more realistic models, that do not permit such an
elementary analysis.

\begin{acknowledgments}
Author thank professor I.I. Lyapilin and professor S.G. Novakshenov for valuable discussions.
The support by  Grant 04-02-96068 RFBR URAL 2004 is acknowledged.
\end{acknowledgments}


\begin{thebibliography}{99}


\bibitem{krum}
Elliott R. J., Krumhansl J. A. and Leath P. L., Rev. Mod. Phys.{\bf 46}, 465 (1974)

\bibitem{hori}
Hori M. \and Yonezawa F. J. Phys. C: Solid State Phys. {\bf 10}, 229 (1976)

\bibitem{GotzePhil}
G\"otze W. Phil. Mag. B {\bf 43}, 219 (1981)

\bibitem{giamarchi}
Chitra R., Giamarchi T. and Le Doussal P. Phys. Rev. Lett. {\bf 80}, 3827 (1998)

\bibitem{GoldFermi}
Gold A., G\"otze W. Phys. Rev. B {\bf 33}, 2495 (1986)

\bibitem{GoldBose}
Gold A. Phys. Rev. A {\bf 33}, 652 (1986)

\bibitem{GoldHTS}
Gold A. Physica C {\bf 190}, 483 (1992)

\bibitem{gruner}
G. Gr\"uner. Rev. Mod. Phys. {\bf 66},1 (1994)

\bibitem{das}
Das D. and Doniach S. Phys. Rev. Lett. {\bf 57}, 1440 (1998), and references cited therein.

\bibitem{fukuyama}
Fukuyama H., Lee P. A. Phys. Rev. B {\bf 17}, 535 (1978)

\bibitem{hines}
Hines D.F., Frankel N.E. Phys. Rev. B {\bf 20}, 972 (1979)

\bibitem{zeitschrift}
Gold A.  Z. Phys. B {\bf 83}, 429 (1991)

\bibitem{ambegaokar}
Ambegaokar V., Halperin B.I., Nelson D. and Siggia E. Rhys. Rev. B {\bf 21}, 1806 (1980

\bibitem{huber}
Huber D.L. Phys. Rev. B {\bf 26}, 3758 (1982)

\bibitem{tanatar}
Davoudi B., Stepparola E., Tanatar B. and Tosi M.P. Phys. Rev. B {\bf 63}, 104505 (2001)

\bibitem{fisher}
Fisher D.S. Phys. Rev. B {\bf 31}, 1396 (1985)

\bibitem{liquids}
G\"otze W., Sj\"ogren L. Rep. Prog. Phys. {\bf 55}, 241 (1992);
Leutheusser E. Phys. Rev. A {\bf 29}, 2765 (1984)

\bibitem{schick}
Schick M. Phys. Rev. A {\bf 3}, 1067 (1971)

\bibitem{nelson}
Nelson D. Phys. Rev. Lett. {\bf 60}, 1973 (1988), and references cited therein.

\end{thebibliography}
\end{document}